\documentclass[fleqn,10pt]{wlscirep}

\usepackage[utf8]{inputenc}
\usepackage[T1]{fontenc}
\usepackage{microtype,newpxtext,newpxmath} 
\usepackage{mathtools,gensymb,siunitx,eurosym,url,soul} 
\usepackage[font=footnotesize,textfont=footnotesize]{subcaption}
\usepackage[font=small,textfont=small]{caption}
\usepackage{longtable,array,booktabs,mdwlist}
\usepackage[colorinlistoftodos,prependcaption,textsize=tiny]{todonotes}
\usepackage{comment,todonotes,lineno,acronym}
\usepackage{xcolor,color}
\usepackage{graphicx}
\usepackage[normalem]{ulem}

\setkeys{Gin}{draft=false} 
\graphicspath{{figures/} {./}}
\makeatletter
\def\ifdraft{\ifdim\overfullrule>\z@
  \expandafter\@firstoftwo\else\expandafter\@secondoftwo\fi}
\makeatother

\acrodef{ADC}[ADC]{Analog to Digital Converter}
\acrodef{ADM}[ADM]{Asynchronous Delta Modulator}
\acrodef{ADEXP}[AdExp-I\&F]{Adaptive-Exponential Integrate and Fire}
\acrodef{AER}[AER]{Address-Event Representation}
\acrodef{AEX}[AEX]{AER EXtension board}
\acrodef{AE}[AE]{Address-Event}
\acrodef{AFM}[AFM]{Atomic Force Microscope}
\acrodef{AGC}[AGC]{Automatic Gain Control}
\acrodef{AMDA}[AMDA]{AER Motherboard with D/A converters}
\acrodef{ANN}[ANN]{Artificial Neural Network}
\acrodef{API}[API]{Application Programming Interface}
\acrodef{ARM}[ARM]{Advanced RISC Machine}
\acrodef{ASIC}[ASIC]{Application Specific Integrated Circuit}
\acrodef{AdExp}[AdExp-IF]{Adaptive Exponential Integrate-and-Fire}
\acrodef{BCM}[BMC]{Bienenstock-Cooper-Munro}
\acrodef{BD}[BD]{Bundled Data}
\acrodef{BEOL}[BEOL]{Back-end of Line}
\acrodef{BG}[BG]{Bias Generator}
\acrodef{BMI}[BMI]{Brain-Machince Interface}
\acrodef{BTB}[BTB]{band-to-band tunnelling}
\acrodef{CAD}[CAD]{Computer Aided Design}
\acrodef{CAM}[CAM]{Content Addressable Memory}
\acrodef{CAVIAR}[CAVIAR]{Convolution AER Vision Architecture for Real-Time}
\acrodef{CA}[CA]{Cortical Automaton}
\acrodef{CCN}[CCN]{Cooperative and Competitive Network}
\acrodef{CDR}[CDR]{Clock-Data Recovery}
\acrodef{CFC}[CFC]{Current to Frequency Converter}
\acrodef{CHP}[CHP]{Communicating Hardware Processes}
\acrodef{CMIM}[CMIM]{Metal-insulator-metal Capacitor}
\acrodef{CML}[CML]{Current Mode Logic}
\acrodef{CMOL}[CMOL]{Hybrid CMOS nanoelectronic circuits}
\acrodef{CMOS}[CMOS]{Complementary Metal-Oxide-Semiconductor}
\acrodef{CNN}[CCN]{Convolutional Neural Network}
\acrodef{COTS}[COTS]{Commercial Off-The-Shelf}
\acrodef{CPG}[CPG]{Central Pattern Generator}
\acrodef{CPLD}[CPLD]{Complex Programmable Logic Device}
\acrodef{CPU}[CPU]{Central Processing Unit}
\acrodef{CSM}[CSM]{Cortical State Machine}
\acrodef{CSP}[CSP]{Constraint Satisfaction Problem}
\acrodef{CTXCTL}[CTXCTL]{Cortex Control}
\acrodef{CV}[CV]{Coefficient of Variation}
\acrodef{DAC}[DAC]{Digital to Analog Converter}
\acrodef{DAS}[DAS]{Dynamic Auditory Sensor}
\acrodef{DAVIS}[DAVIS]{Dynamic and Active Pixel Vision Sensor}
\acrodef{DBN}[DBN]{Deep Belief Network}
\acrodef{DFA}[DFA]{Deterministic Finite Automaton}
\acrodef{DIBL}[DIBL]{drain-induced-barrier-lowering}
\acrodef{DI}[DI]{delay insensitive}
\acrodef{DMA}[DMA]{Direct Memory Access}
\acrodef{DNF}[DNF]{Dynamic Neural Field}
\acrodef{DNN}[DNN]{Deep Neural Network}
\acrodef{DOF}[DOF]{Degrees of Freedom}
\acrodef{DPE}[DPE]{Dynamic Parameter Estimation}
\acrodef{DPI}[DPI]{Differential Pair Integrator}
\acrodef{DRRZ}[DR-RZ]{Dual-Rail Return-to-Zero}
\acrodef{DRAM}[DRAM]{Dynamic Random Access Memory}
\acrodef{DR}[DR]{Dual Rail}
\acrodef{DSP}[DSP]{Digital Signal Processor}
\acrodef{DVS}[DVS]{Dynamic Vision Sensor}
\acrodef{DYNAP}[DYNAP]{Dynamic Neuromorphic Asynchronous Processor}
\acrodef{EBL}[EBL]{Electron Beam Lithography}
\acrodef{EDVAC}[EDVAC]{Electronic Discrete Variable Automatic Computer}
\acrodef{EEG}[EEG]{electroencephalography}
\acrodef{EIN}[EIN]{Excitatory-Inhibitory Network}
\acrodef{EM}[EM]{Expectation Maximization}
\acrodef{EPSC}[EPSC]{Excitatory Post-Synaptic Current}
\acrodef{EPSP}[EPSP]{Excitatory Post-Synaptic Potential}
\acrodef{EZ}[EZ]{Epileptogenic Zone}
\acrodef{FDSOI}[FDSOI]{Fully-Depleted Silicon on Insulator}
\acrodef{FET}[FET]{Field-Effect Transistor}
\acrodef{FFT}[FFT]{Fast Fourier Transform}
\acrodef{FI}[F-I]{Frequency-Current}
\acrodef{FPGA}[FPGA]{Field Programmable Gate Array}
\acrodef{FR}[FR]{Fast Ripple}
\acrodef{FSA}[FSA]{Finite State Automaton}
\acrodef{FSM}[FSM]{Finite State Machine}
\acrodef{GIDL}[GIDL]{gate-induced-drain-leakage}
\acrodef{GOPS}[GOPS]{Giga-Operations per Second}
\acrodef{GPU}[GPU]{Graphical Processing Unit}
\acrodef{GUI}[GUI]{Graphical User Interface}
\acrodef{HAL}[HAL]{Hardware Abstraction Layer}
\acrodef{HFO}[HFO]{High Frequency Oscillation}
\acrodef{HH}[H\&H]{Hodgkin \& Huxley}
\acrodef{HMM}[HMM]{Hidden Markov Model}
\acrodef{HRS}[HRS]{High-Resistive State}
\acrodef{HR}[HR]{Human Readable}
\acrodef{HSE}[HSE]{Handshaking Expansion}
\acrodef{HW}[HW]{Hardware}
\acrodef{ICT}[ICT]{Information and Communication Technology}
\acrodef{IC}[IC]{Integrated Circuit}
\acrodef{IEEG}[iEEG]{intracranial electroencephalography}
\acrodef{IF2DWTA}[IF2DWTA]{Integrate \& Fire 2--Dimensional WTA}
\acrodef{IFSLWTA}[IFSLWTA]{Integrate \& Fire Stop Learning WTA}
\acrodef{IF}[I\&F]{Integrate-and-Fire}
\acrodef{IMU}[IMU]{Inertial Measurement Unit}
\acrodef{INCF}[INCF]{International Neuroinformatics Coordinating Facility}
\acrodef{INI}[INI]{Institute of Neuroinformatics}
\acrodef{IO}[I/O]{Input/Output}
\acrodef{IPSC}[IPSC]{Inhibitory Post-Synaptic Current}
\acrodef{IPSP}[IPSP]{Inhibitory Post-Synaptic Potential}
\acrodef{IP}[IP]{Intellectual Property}
\acrodef{ISI}[ISI]{Inter-Spike Interval}
\acrodef{IoT}[IoT]{Internet of Things}
\acrodef{JFLAP}[JFLAP]{Java - Formal Languages and Automata Package}
\acrodef{LEDR}[LEDR]{Level-Encoded Dual-Rail}
\acrodef{LFP}[LFP]{Local Field Potential}
\acrodef{LLC}[LLC]{Low Leakage Cell}
\acrodef{LNA}[LNA]{Low-Noise Amplifier}
\acrodef{LPF}[LPF]{Low Pass Filter}
\acrodef{LRS}[LRS]{Low-Resistive State}
\acrodef{LSM}[LSM]{Liquid State Machine}
\acrodef{LTD}[LTD]{Long Term Depression}
\acrodef{LTI}[LTI]{Linear Time-Invariant}
\acrodef{LTP}[LTP]{Long Term Potentiation}
\acrodef{LTU}[LTU]{Linear Threshold Unit}
\acrodef{LUT}[LUT]{Look-Up Table}
\acrodef{LVDS}[LVDS]{Low Voltage Differential Signaling}
\acrodef{MCMC}[MCMC]{Markov-Chain Monte Carlo}
\acrodef{MEMS}[MEMS]{Micro Electro Mechanical System}
\acrodef{MFR}[MFR]{Mean Firing Rate}
\acrodef{MIM}[MIM]{Metal Insulator Metal}
\acrodef{MLP}[MLP]{Multilayer Perceptron}
\acrodef{MOSCAP}[MOSCAP]{Metal Oxide Semiconductor Capacitor}
\acrodef{MOSFET}[MOSFET]{Metal Oxide Semiconductor Field-Effect Transistor}
\acrodef{MOS}[MOS]{Metal Oxide Semiconductor}
\acrodef{MRI}[MRI]{Magnetic Resonance Imaging}
\acrodef{NDFSM}[NDFSM]{Non-deterministic Finite State Machine} 
\acrodef{ND}[ND]{Noise-Driven}
\acrodef{NEF}[NEF]{Neural Engineering Framework}
\acrodef{NHML}[NHML]{Neuromorphic Hardware Mark-up Language}
\acrodef{NIL}[NIL]{Nano-Imprint Lithography}
\acrodef{NMDA}[NMDA]{N-Methyl-D-Aspartate}
\acrodef{NME}[NE]{Neuromorphic Engineering}
\acrodef{NN}[NN]{Neural Network}
\acrodef{NRZ}[NRZ]{Non-Return-to-Zero}
\acrodef{NSM}[NSM]{Neural State Machine}
\acrodef{OR}[OR]{Operating Room}
\acrodef{OTA}[OTA]{Operational Transconductance Amplifier}
\acrodef{PCB}[PCB]{Printed Circuit Board}
\acrodef{PCHB}[PCHB]{Pre-Charge Half-Buffer}
\acrodef{PCM}[PCM]{Phase Change Memory}
\acrodef{PE}[PE]{Phase Encoding}
\acrodef{PFA}[PFA]{Probabilistic Finite Automaton}
\acrodef{PFC}[PFC]{prefrontal cortex}
\acrodef{PFM}[PFM]{Pulse Frequency Modulation}
\acrodef{PR}[PR]{Production Rule}
\acrodef{PSC}[PSC]{Post-Synaptic Current}
\acrodef{PSP}[PSP]{Post-Synaptic Potential}
\acrodef{PSTH}[PSTH]{Peri-Stimulus Time Histogram}
\acrodef{QDI}[QDI]{Quasi Delay Insensitive}
\acrodef{RA}[RA]{Resected Area}
\acrodef{RAM}[RAM]{Random Access Memory}
\acrodef{RDF}[RDF]{random dopant fluctuation}
\acrodef{RELU}[ReLu]{Rectified Linear Unit}
\acrodef{RLS}[RLS]{Recursive Least-Squares}
\acrodef{RMSE}[RMSE]{Root Mean Squared-Error}
\acrodef{RMS}[RMS]{Root Mean Squared}
\acrodef{RNN}[RNN]{Recurrent Neural Networks}
\acrodef{ROLLS}[ROLLS]{Reconfigurable On-Line Learning Spiking}
\acrodef{RRAM}[R-RAM]{Resistive Random Access Memory}
\acrodef{R}[R]{Ripples}
\acrodef{SAC}[SAC]{Selective Attention Chip}
\acrodef{SAT}[SAT]{Boolean Satisfiability Problem}
\acrodef{SCX}[SCX]{Silicon CorteX}
\acrodef{SD}[SD]{Signal-Driven}
\acrodef{SEM}[SEM]{Spike-based Expectation Maximization}
\acrodef{SLAM}[SLAM]{Simultaneous Localization and Mapping}
\acrodef{SNN}[SNN]{Spiking Neural Network}
\acrodef{SNR}[SNR]{Signal to Noise Ratio}
\acrodef{SOC}[SOC]{System-On-Chip}
\acrodef{SOI}[SOI]{Silicon on Insulator}
\acrodef{SOZ}[SOZ]{Seizure Onset Zone}
\acrodef{SP}[SP]{Separation Property}
\acrodef{SRAM}[SRAM]{Static Random Access Memory}
\acrodef{STDP}[STDP]{Spike-Timing Dependent Plasticity}
\acrodef{STD}[STD]{Short-Term Depression}
\acrodef{STP}[STP]{Short-Term Plasticity}
\acrodef{STT-MRAM}[STT-MRAM]{Spin-Transfer Torque Magnetic Random Access Memory}
\acrodef{STT}[STT]{Spin-Transfer Torque}
\acrodef{SW}[SW]{Software}
\acrodef{TCAM}[TCAM]{Ternary Content-Addressable Memory}
\acrodef{TLE}[TLE]{Temporal Lobe Epilepsy}
\acrodef{TFT}[TFT]{Thin Film Transistor}
\acrodef{USB}[USB]{Universal Serial Bus}
\acrodef{VHDL}[VHDL]{VHSIC Hardware Description Language}
\acrodef{VLSI}[VLSI]{Very Large Scale Integration}
\acrodef{VOR}[VOR]{Vestibulo-Ocular Reflex}
\acrodef{WCST}[WCST]{Wisconsin Card Sorting Test}
\acrodef{WTA}[WTA]{Winner-Take-All}
\acrodef{XML}[XML]{eXtensible Mark-up Language}
\acrodef{CTXCTL}[CTXCTL]{CortexControl}
\acrodef{divmod3}[DIVMOD3]{divisibility of a number by three}
\acrodef{hWTA}[hWTA]{hard Winner-Take-All}
\acrodef{sWTA}[sWTA]{soft Winner-Take-All}

\title{A Spiking Neural Network (SNN) for detecting High Frequency Oscillations (HFOs) in the intraoperative ECoG}
\author[1,2]{Karla Burelo}
\author[1,2]{Mohammadali Sharifshazileh}
\author[3,4]{Niklaus Krayenbühl}
\author[3,4]{Georgia Ramantani}
\author[1]{Giacomo Indiveri}
\author[2,4*]{Johannes Sarnthein}

\affil[1]{Institute of Neuroinformatics, University of Zurich and ETH Zürich, 8057 Zurich, Switzerland}
\affil[2]{Klinik für Neurochirurgie, UniversitätsSpital und Universität Zürich, 8091 Zurich, Switzerland}
\affil[3]{UniversitätsKinderspital,  8032 Zurich, Switzerland}
\affil[4]{Klinisches Neurowissenschaften Zentrum, Universität Zürich, Switzerland}
\affil[*]{johannes.sarnthein@usz.ch}

\begin{abstract}
To achieve seizure freedom, epilepsy surgery requires the complete resection of the epileptogenic brain tissue. In intraoperative ECoG recordings, high frequency oscillations (HFOs) generated by epileptogenic tissue can be used to tailor the resection margin. However, automatic detection of HFOs in real-time remains an open challenge. Here we present a spiking neural network (SNN) for automatic HFO detection that is optimally suited for neuromorphic hardware implementation. 

We trained the SNN to detect HFO signals measured from intraoperative ECoG on-line, using an independently labeled dataset. We targeted the detection of HFOs in the fast ripple frequency range (250-500 Hz) and compared the network results with the labeled HFO data. 

We endowed the SNN with a novel artifact rejection mechanism to suppress sharp transients and demonstrate its effectiveness on the ECoG dataset. The HFO rates (median 6.6 HFO/min in pre-resection recordings) detected by this SNN are comparable to those published in the dataset (58 min, 16 recordings). The postsurgical seizure outcome was ‘predicted’ with 100\% accuracy for all 8 patients.

These results provide a further step towards the construction of a real-time portable battery-operated HFO detection system that can be used during epilepsy surgery to guide the resection of the epileptogenic zone.

\end{abstract}

\begin{document}

\flushbottom
\maketitle

\newpage

\section{Introduction}
\label{sec:introduction}
Among patients with epilepsy, one-third have seizures that cannot be controlled by medication~\cite{Ryvlin_etal14}. Selected patients with drug-resistant focal epilepsy may benefit from epilepsy surgery to achieve seizure freedom. The efficacy of epilepsy surgery requires the complete resection of the epileptogenic brain tissue~\cite{Jette_etal14}. 
Intraoperative electrocorticography (ECoG) can be performed during surgery to optimize the delineation of the epileptogenic area against healthy brain tissue by taking into account interictal spike patterns~\cite{Lesko_etal20,Demuru_etal20,Grewal_etal19,Groppel_etal19}.
This so called ``tailoring'' may guide surgical decisions, but the value of interictal spikes as an epilepsy biomarker in this context is under debate.

Interictal high frequency oscillations (HFO > 80\,Hz), particularly in the fast ripple band (250-500 Hz), have recently been established as biomarkers for the intraoperative delineation of the epileptogenic zone~\cite{Boran_etal19,Fedele_etal17b,Fedele_etal16,Weiss_etal18,Vant-klooster_etal15a,Vant-klooster_etal17,Vant-klooster_etal15b,Wang_etal20}.
HFO detection faces the challenge of low signal-to-noise ratio, which can be improved by high density electrodes~\cite{Boran_etal19,Zweiphenning_etal20} and by low-noise amplification~\cite{Fedele_etal17b}.
As a further challenge, a clinically relevant HFO must be distinguished from the electrical artifacts induced by the standard intraoperative devices or any other spurious oscillation in the fast ripple band.
To achieve clinical relevance, an HFO should 1) be defined prospectively and 2) be validated against postsurgical seizure freedom~\cite{Fedele_etal19}. 
While there are many automated detection schemes that define HFOs prospectively, only few validated the detected HFOs against postsurgical seizure freedom~\cite{Boran_etal19,Fedele_etal17b,Fedele_etal16,Weiss_etal18}.
These detectors however, require further offline processing of the pre-recorded signal to apply an automatic or semiautomatic artifact rejection stage to eliminate events wrongly classified as HFO. 
This offline processing requirement limits the possibility to perform real-time detection of HFOs during the time-span available within the constraints of the surgery.

In order to implement the clinical use of HFOs in ECoG, their value has yet to be confirmed in a prospective clinical trial. A first small study is expected to report results on the non-inferiority of HFOs compared to spikes~\cite{Vant-klooster_etal15a}. Evidence for superiority of HFOs over spikes requires a multi-center clinical trial with larger numbers of patients, which will need fast, reliable, and unsupervised automatic HFO detection to provide a standardized definition of clinically relevant HFOs.

Here, we simulated a spiking neural network (SNN) for HFO detection in the intraoperative ECoG.
This work builds on a previously validated SNN for HFO detection in the intracranial EEG (iEEG)~\cite{Sharifshazileh_etal20, Sharifshazileh_etal19}, and extends it by introducing a novel artifact rejection mechanism to reject fast transient artifacts, and by validating it on intraoperative electrocorticography recordings (ECoG).
As a computational principle, the SNN emulates the spiking of neurons in small networks~\cite{Roy_etal19}  so that they can be implemented in low-power and compact neuromorphic hardware that perform real-time computation~\cite{Chicca_etal14}.

We applied our SNN to a published dataset of pre-recorded ECoG, where HFOs were detected offline using an automated HFO detector (Spectrum detector)~\cite{Fedele_etal16,Burnos_etal14} and validated against postoperative seizure freedom~\cite{Boran_etal19}.
In this benchmark testing, we were able to correctly predict the postoperative seizure outcome in all 8 patients. 
This is a further step towards an SNN that may be implemented in a neuromorphic device~\cite{Sharifshazileh_etal20} for standardized and real-time HFO detection during epilepsy surgery.

\section{Methods}
\label{sec:methods}

\subsection{Patients}
We retrospectively included pediatric and adult patients (median age 18.5 y, range [12, 33 ] y) who 1) underwent epilepsy surgery in our institutions, 2) where the resection was guided by intraoperative high-density ECoG (hd-ECoG), 3) post-resection hd-ECoG was available, and 4) the follow-up-duration after surgery was $\geq$ 12 months.

The patients were were followed-up at the outpatient clinics of Neurosurgery, University Hospital Zurich, and Epileptology, University Childrens Hospital Zurich, 3, 6, 12, 24 months after surgery according to our institutional protocol.
Postsurgical seizure outcome was determined according to the International League Against Epilepsy (ILAE) scale and consecutively classified in two categories: seizure freedom (ILAE 1) and seizure recurrence (ILAE 2-5). 

\subsection{Ethical considerations}
The collection of personal patient data and their analysis were
approved and performed conform to the guidelines and regulations
of the local research ethics committee (Kantonale Ethikkommission Zürich KEK-ZH-Nr. 2019-01977), who waived the collection of patients’ written informed consent.

\subsection{Anesthesia management}
According to our standard protocol for neurosurgical interventions, anesthesia was induced with intravenous application of Propofol (1.5–2 mg/kg) and Fentanyl (2–3  $\mu$g/kg). 
Intratracheal intubation was facilitated by Atracurium (0.5 mg/kg). Anesthesia was maintained with Propofol (5–10 mg/kg/h) and Remifentanil (0.1–2 $\mu$g/kg/min). 
Twenty minutes before ECoG recording, Propofol was ceased and anesthesia was sustained with Sevoflurane (MAC<0.5). 

\subsection{ECoG recordings}
Intraoperative ECoG was recorded with high-density subdural grid electrodes (hd-ECoG, AdTech Medical, contact exposure diameter 2.3\,mm, inter-electrode distance 5\,mm). We used a needle electrode placed in the dura as electrical reference. We collected ECoG data with a Nicolet recording device (Nicolet\textsuperscript{\tiny\textregistered} CSeries amplifier: Natus Medical Incorporated, 16-bit ADC, Pleasanton, PA, USA; sampling rate 2\,kHz, 1-800\,Hz passband). 
All ECoG data was re-referenced to a bipolar montage along the length of the grid. Intervals with visible artifacts and channels affected by continuous interference or not recording from brain tissue were excluded from further analysis. 
The placement of subdural electrodes for intraoperative ECoG was guided solely by the clinical question. Only interictal epileptiform discharges (and not HFO) were considered for the intraoperative delineation of the epileptogenic zone and thus for tailoring the resection.
In this study, we first analysed the ECoG recorded from a location over the volume to be resected and its margins (pre-resection ECoG), and then the post-resection ECoG that was recorded at the resection margins. 
The neurosurgeon and the neurologist in charge of the patient carefully documented the ECoG electrode localization with respect to the resected volume. 
This data is publicly available as described earlier~\cite{Boran_etal19} and can be found at~\url{https://gin.g-node.org/USZ_NCH/Intraoperative_ECoG_HFO}.

\subsection{HFO detection with the Spectrum detector}
The Spectrum detector has been described in detail in previous publications~\cite{Boran_etal19,Fedele_etal17b,Fedele_etal16,Burnos_etal14}. 
In brief, the detector has three stages. Stage I determined a baseline amplitude threshold in time intervals with high Stockwell entropy (low oscillatory activity). Events exceeding the threshold were marked as events of interest (EoI). In Stage II, we selected all EoI that exhibited a high frequency peak isolated from low frequency activity in the time–frequency space~\cite{Burnos_etal14}. The number of EoI was further reduced in Stage III, where artifacts with multichannel spread were rejected, since HFO are spatially confined in a small patch of cortical tissue~\cite{Burnos_etal16b}. This prospective definition of a clinically relevant HFO has been shown to predict postsurgical seizure outcome with high accuracy~\cite{Boran_etal19,Fedele_etal17b,Fedele_etal16}. 
Following these steps of automated HFO detection, an observer inspected the events in wideband and filtered in the HFO band to reject further artifacts.

\subsection{HFO detection with the SNN}
\begin{figure}
\centering\includegraphics[width=\linewidth]{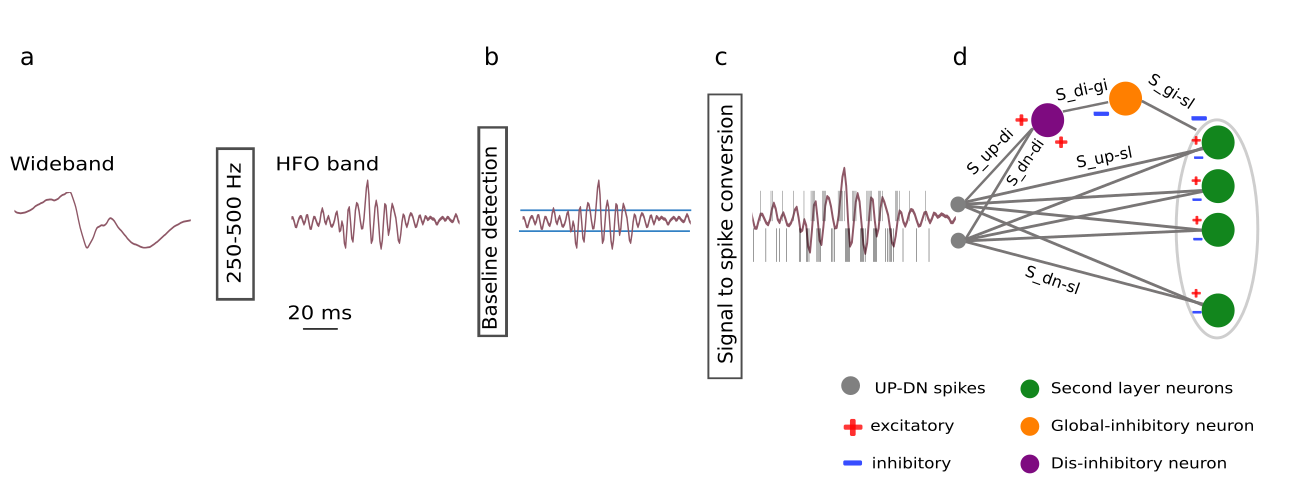}
  \caption{HFO detection scheme. (a) The wideband ECoG is filtered in the HFO frequency band (250-500 Hz) (b) In the baseline detection stage, the background noise of the signal is used to set the signal-to-spike threshold. (c) The signal-to-spike conversion algorithm converts the analog signal into two streams of digital outputs: UP and DN spikes. (d) The SNN architecture for HFO detection and artifact rejection consists of input neurons (grey) receiving the input UP-DN spikes. These inputs project to a second layer of neurons (green) and to a dis-inhibitory neuron (purple). This neuron projects inhibitory synapses to an global-inhibitory neuron (orange), which is continuously inhibiting the second layer neurons. The synapses of the projections are excitatory (positve, red) or inhibitory (negative, blue).}
\label{fig:Pipeline}
\end{figure}

As a first step in the HFO detection pipeline, the wideband ECoG recording was filtered in the 250-500\,Hz fast ripple band (Fig.~\ref{fig:Pipeline}a). Since Butterworth filters showed to be a good approximation of the Tow-Thomas architectures for the hardware SNN filters~\cite{Sharifshazileh_etal19,Sharifshazileh_etal20,Fleischer_Tow73}, we used a 2nd order filter of this model and the SciPy python package to simulate the filtering stage \cite{Butterworth_etal30, Selesnick_etal98, Virtanen_etal20}.
In the filtered signal, we defined a baseline amplitude that has to be exceeded by a putative HFO event (Fig.~\ref{fig:Pipeline}b). Following the algorithm implemented in the hardware SNN~\cite{Sharifshazileh_etal20}, we selected a 1\,s time window, stored the maximum signal amplitudes of consecutive non-overlapping time windows of 50\,ms, and took the mean of the lowest quartile as the baseline amplitude.

The filtered signal was then converted to spikes using a delta conversion scheme (Fig.~\ref{fig:Pipeline}c) that is inspired by the analog delta modulator (ADM)~\cite{Yang_etal15,Corradi_Indiveri15}. The ADM integrates an error signal which follows the input until it increases above (decreases below) the upper (lower) threshold. Upon the crossover, an UP (DN) spike is generated and the error will reset to zero. The error will remain zero for a refractory period after which, it will continue tracking the input again until the next threshold crossover, hence encoding the input into UP-DN spike trains. This threshold was set at 50\% of the pre-recorded ECoG baseline amplitude. To approximate the asynchronous conversion of the signal in the ADM, we over-sampled the input signal at 35\,kHz and we set a refractory period (300\,$\mu$S) to simulate the delay of the ADM after each spike. Note that over-sampling period should be much smaller than the refractory period for accurate data conversion.

The HFO detection stage of the network (the core SNN) consists of input neurons receiving the input UP-DN spikes and a second layer of neurons (Fig.~\ref{fig:Pipeline}d). The projections to the second layer neurons are excitatory for UP spikes and inhibitory for DN spikes.
We used the Python SNN simulator Brian2~\cite{Goodman_Brette08}, the custom toolbox teili~\cite{Milde_etal18}, and the parameters in Table~\ref{tab:SNN_parameters} to simulate an SNN that matches the behavior of the neuromorphic circuits of the hardware SNN~\cite{Sharifshazileh_etal20}.
The software simulations take into account the neuromorphic circuit properties. As the circuits are based on a ``current-mode'' design, we represented all the relevant state variables (such as the neuron membrane potential) as currents (e.g., $I_{mem}$).

Spikes in the second layer neurons were used to mark an HFO.
Any spike within a 15\,ms window indicated an HFO, where consecutive windows containing spikes were concatenated to indicate the same HFO. There were no further steps required for artifact rejection.
HFO detection was performed independently for each channel of the pre- and post-resection ECoG recordings. 

\subsection{Prediction of seizure outcome using residual HFO}
The output of the HFO detection was compared to the postsurgical seizure outcome in each patient. For each patient, we calculated the HFO rate in each electrode channel of the pre- and post-resection recordings by dividing the number of HFOs detected in the channel by the duration of the recording. We compared the HFO rate between pre- and post- resection recordings by only selecting the recording channel that had the highest HFO rate (Table~\ref{table:patient_characteristics}), since the presence of a single channel with residual HFO has shown to predict seizure recurrence~\cite{Boran_etal19,Fedele_etal17b,Fedele_etal16,Weiss_etal18,Vant-klooster_etal15a,Vant-klooster_etal17,Vant-klooster_etal15b}.
Channels with HFO rates of $\geq$ 1 HFO/min in the last post-resection ECoG were defined as having residual HFOs~\cite{Boran_etal19,Fedele_etal16}.

To quantify the predictive value of HFO with respect to seizure outcome in each patient, we retrospectively ``predicted'' seizure freedom (ILAE 1) in patients with post-resection HFO rates of $<$ 1 HFO/min and recurrent seizures (ILAE 2-6) in patients with post-resection HFO rates of $\geq$ 1 HFO/min (Table~\ref{table:patient_characteristics}).
We divided the patients into four groups. True Positive (TP): residual HFO, seizure recurrence (ILAE $>$1) correctly predicted; True Negative (TN): no residual HFO, seizure freedom (ILAE 1) correctly predicted; False Positive (FP): residual HFO, seizure freedom falsely predicted; False Negative (FN): no residual HFO, seizure recurrence falsely predicted. 
The positive predictive value was calculated as PPV=TP/(TP+FP), negative predictive value as NPV=TN/(TN+FN), sensitivity as Sens=TP/(TP+FN), specificity as Spec=TN/(TN+FP), and accuracy as ACC=(TP+TN)/(TP+TN+FP+FN). We calculated the 95\% confidence interval (CI) on the basis of the binomial distribution.

\begin{table}
\tiny
\caption{Synapse parameters of the SNN detector. A connection between two neurons (Figure~\ref{fig:Pipeline}d) is characterized by the positive (excitatory, exc) or negative (inhibitory, inh) current in fA and the time constant.}
\label{tab:SNN_parameters}
\renewcommand\arraystretch{1.2}
\centering
\resizebox{\textwidth}{!}{\begin{tabular}{@{}
>{\arraybackslash}p{0.2\textwidth}
>{\centering\arraybackslash}p{0.02\textwidth} >{\centering\arraybackslash}p{0.06\textwidth} >{\centering\arraybackslash}p{0.01\textwidth}
>{\centering\arraybackslash}p{0.09\textwidth}@{}}
\toprule
\textbf{Connection} & \textbf{Name} & \textbf{Connection strength (fA)} & \textbf{Polarity} &\textbf{Time constant $\tau$ (ms)}\\
\hline
Input UP spikes to second layer  &$S_{up-sl}$ & [7 14] & exc & [3 6]\\
Input DN spikes to second layer  &$S_{dn-sl}$ & [7 14] & inh & $S_{up-sl}$-[0.1 1]\\
Input UP spikes to dis-inhibitory neuron &$S_{up-di}$ & 21 & exc & 5\\
Input DN spikes to dis-inhibitory neuron &$S_{dn-di}$ & 21 & exc & 5\\
Dis-inhibitory neuron to global-inhibitory neuron &$S_{di-gi}$ & 17.5 &inh & 20\\
Poisson to global-inhibitory neuron &$S_{poiss-gi}$ & - & exc & 5\\
global-inhibitory neuron to second layer &$S_{gi-sl}$ & 24.5 & inh & 5\\
\bottomrule
\end{tabular}}
\vspace{-0.31cm}
\end{table}

\begin{figure}
\centering\includegraphics[width=0.8\linewidth]{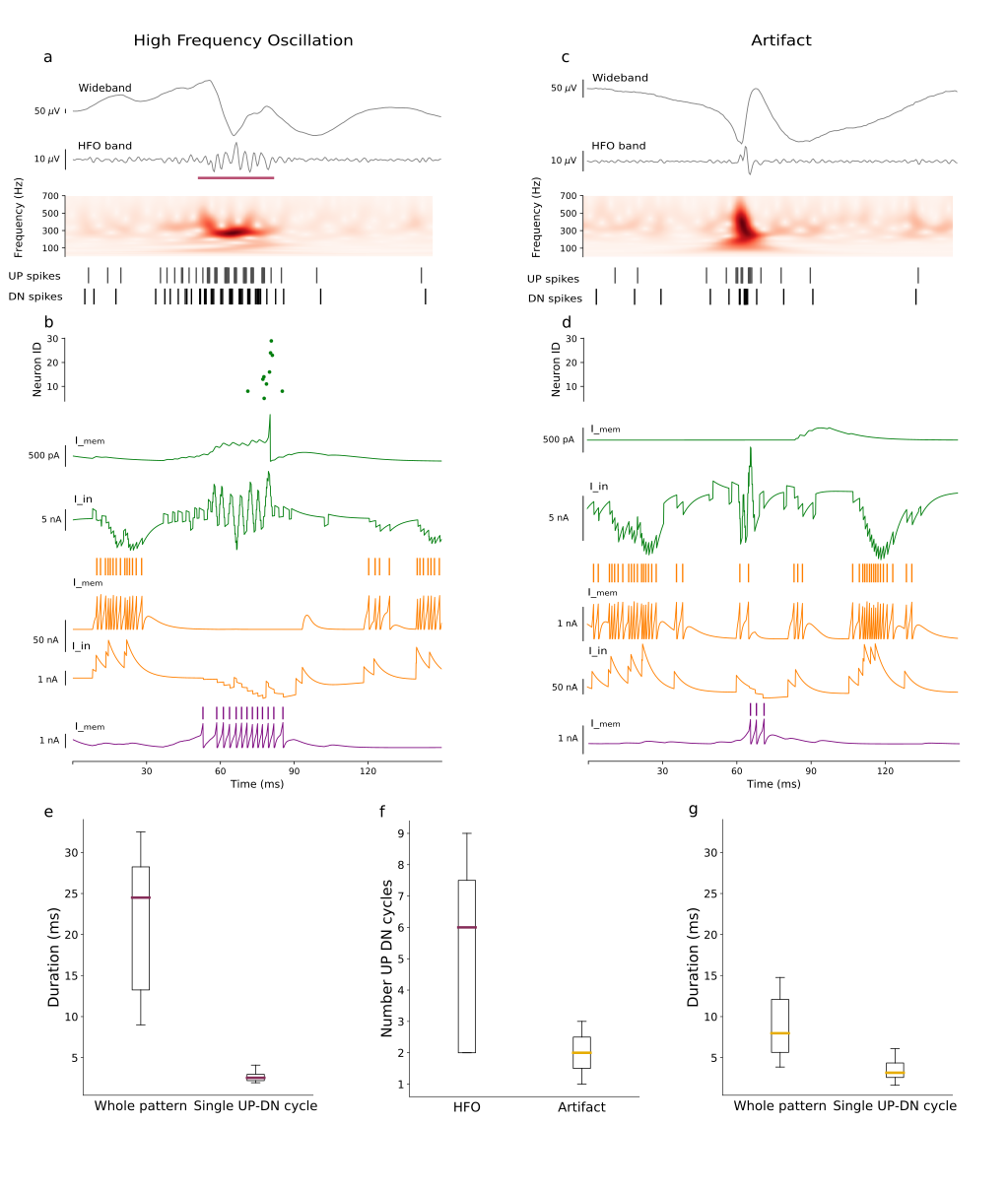}
  \caption{The SNN distinguishes between an HFO and an artifact (fast transient) in the ECoG. 
  (a,c) ECoG signal and input spike train to the SNN. Consecutive UP-DN spike bursts in the train define an UP-DN cycle. More cycles occurred during the HFO. 
  (b,d) Activity of the neurons in the SNN. 
  (b) During an HFO, the input spike train excited the dis-inhibitory neuron (membrane potential $I_{mem}$ in purple), which suppressed the global-inhibitory neuron that became silent (input current $I_{in}$ and $I_{mem}$ in orange). A neuron of the core SNN (input current $I_{in}$ and membrane potential $I_{mem}$ in green) integrated the input spike trains and produced an output spike. Several second layer neurons of the core SNN responded to the HFO (green dots).
  (d) During a short fast transient in the ECoG, the excitation of the dis-inhibitory neuron was so short (purple trace) that it did not silence the global-inhibitory neuron (orange traces), which in turn continued to inhibit the second layer neuron (green traces) and prevented the generation of an output spike. Neither this nor any other neuron of the second layer $I_{mem}$ (green) reached the spiking threshold hence, the raster plot remained empty. 
  (e,f,g) Population characteristics of the UP-DN input spike trains entering the second layer neurons and the dis-inhibitory neuron. 
  (e) The train during an HFO lasted 24\,ms (median), with a single cycle lasting 2.6\,ms (median). 
  (g) The train during an artifact lasted 9\,ms (median), with a single cycle lasting 3.2\,ms (median). 
  (f) The HFOs comprise more UP-DN cycles (median 6 cycles) than the artifacts (median 2 cycles). The cycle characteristics were used to select the parameters of the dis-inhibitory neuron and the global-inhibitory neuron (Table~\ref{tab:SNN_parameters}).}
 \label{fig:artifact_rejection}
\end{figure}


\section{Results}
\label{sec:results}
\subsection{Architecture of the core SNN augmented by the artifact rejection stage}
As a first result, we describe here the architecture of the SNN that is augmented by the artifact rejection stage (Figure~\ref{fig:Pipeline}d). For the core SNN, we used a two-layered feed-forward network of integrate and fire neurons with dynamic synapses~\cite{Sharifshazileh_etal20}. For the first layer of the network, we used two input neurons that projected UP-DN spike trains to the second layer neurons using excitatory and inhibitory synapses, respectively. As for the artifact rejection stage, we implemented a dis-inhibitory mechanism using a dis-inhibitory neuron and an global-inhibitory neuron that constantly suppressed the activity of the second layer neurons. For this purpose, the global-inhibitory neuron was stimulated with a Poisson spike train to generate continuous spikes at 135 Hz. The dis-inhibitory neuron projected inhibitory synapses to the global-inhibitory neuron and received the input UP-DN spikes through excitatory synapses. 

\subsection{Parameters for the artifact rejection stage}
To calibrate the artifact rejection stage, the optimal parameters were found heuristically by analyzing HFOs and sharp transients in the ECoG as follows. 
We composed a training signal of 22 snippets (50\,ms wide) of which 11  contained an HFO that had been marked with the Spectrum detector and published together with the dataset~\cite{Boran_etal19}. For the other 11 snippets, we visually selected fast transients in the wideband signal that the core SNN wrongly classified as HFO. 
The signals were then transformed in UP-DN spike trains (Figure \ref{fig:artifact_rejection}a,c). 
In a spike train, we defined as an ``UP-DN cycle'' a burst of UP spikes followed by a burst of DN spikes.
From the spike trains we quantified 1) the duration of the spike train, 2) the number of cycles and 3) the inter-spike-interval (ISI) within the bursts of a cycle.
We observed that fast transients generated two cycles (median) in the UP-DN spike train while HFOs generated more cycles (median 6 cycles, Figure ~\ref{fig:artifact_rejection}f). 

To inhibit the second layer neurons during a fast transient, the global-inhibitory neuron must suppress their activity for at least one cycle. Conversely, to dis-inhibit the activity of the second layer neurons during an HFO, the dis-inhibition should start as early as possible (i.e. after the first cycle) and should last the total duration of the HFO. In our data set, the median HFO duration was 24\,ms with minimum 9\,ms and the median duration of an artifact was 8\,ms (Fig.~\ref{fig:artifact_rejection}e,g). Although the median duration for fast transients was shorter than for HFOs, the median duration of a single cycle during a fast transient was larger (3.2\,ms) than a single cycle during an HFO (2.6\,ms). Therefore, the suppression of the activity of the second layer neurons during the first cycle resulted in a suppression of HFOs with short duration. However, this design choice did not affect the maximum HFO rates in the benchmarking between our SNN detector and the Spectrum detector (Table~\ref{table:patient_characteristics}).

Since the dis-inhibitory neuron received excitatory inputs from both UP and DN spikes, any activity in the signal could have caused the activation of the dis-inhibitory neuron and, in consequence, the dis-inhibition of the second-layer neurons and an erroneous HFO detection. We avoided this dis-inhibition by using a short synaptic time constant for the connections of the dis-inhibitory neuron. Hence, the dis-inhibitory neuron was activated only during periods of elevated UP-DN spiking as it occurred during a fast transient or an HFO.

\subsection{Example of a detected HFO and a suppressed transient}
The SNN activity differed markedly between detection of an HFO and rejection of a fast transient (Figure~\ref{fig:artifact_rejection} b,d). The UP-DN spike train generated spikes in the membrane potential of the dis-inhibitory neuron (purple trace $I_{mem}$). Note that one UP-DN cycle has passed until the dis-inhibitory neuron responded to the UP-DN inputs. Only after this delay of one cycle, this neuron inhibited the global-inhibitory neuron (orange $I_{in}$ trace).
Note that this trace results from integrating the inputs from the Poisson spike train and the inhibitory inputs from the dis-inhibitory neuron.

During the presence of an HFO in the signal, the global-inhibitory neuron remained silent (\textasciitilde30\,ms, purple bar in Panel\,a). Meanwhile, the neuron in the second layer integrated the UP-DN spikes in its input signal (green $I_{in}$ trace Panel\,b) and accumulated enough evidence in its membrane potential (green $I_{mem}$ trace) to generate a spike in response to the HFO. The raster plot on the top of Panel\,b shows the spikes of other neurons in the second layer that also responded to this HFO.
Detection of an HFO was defined as a spike of at least one second layer neuron.

During the presence of a fast transient in the signal, all the neurons in the second layer remained silent as seen in the raster plot in the top of Panel\,d. Similarly as in the HFO example, once one UP-DN cycle has passed, the dis-inhibitory neuron responded to the UP-DN inputs. However, its activation was not sufficient to suppress the activity of the global-inhibitory neuron. This neuron was active during the presence of the transient (orange traces) and inhibited the second layer neuron. The second layer neuron integrated the UP-DN spikes, which increased the green input current $I_{in}$. However, before it could accumulate enough evidence, the neuron was inhibited by the global-inhibitory neuron which kept this neuron and the whole second layer silent. Thus, the dis-inhibitory neuron was activated only when an HFO was present in the signal; fast transients in the ECoG were suppressed in the SNN and not misclassified as HFO.

\subsection{Residual HFO predict poor seizure outcome}
\begin{figure}
\centering\includegraphics[width=\linewidth]{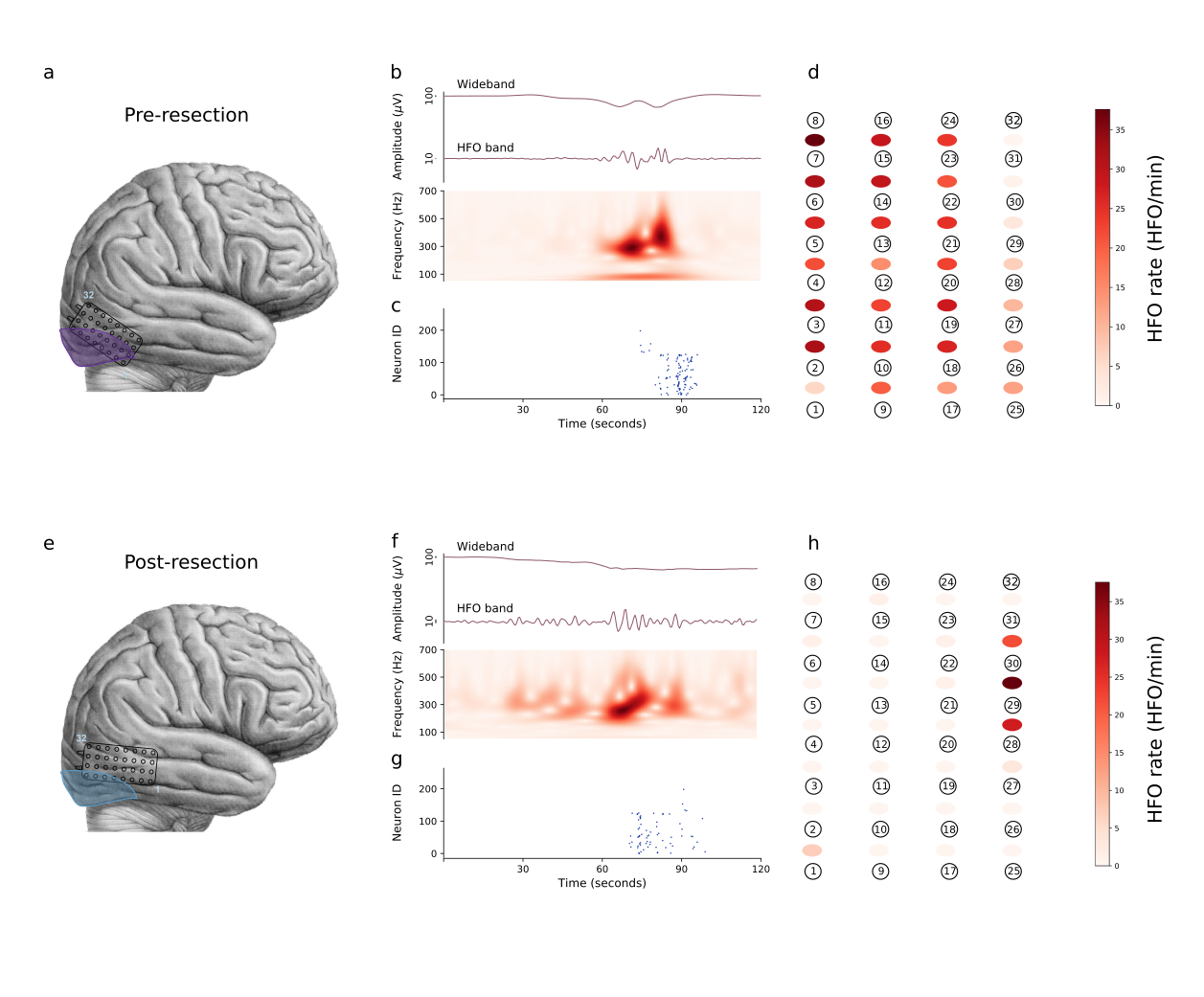}
  \caption{Intraoperative ECoG recording and HFO detection in Patient 6. 
  (a) Position of the high-density grid electrode before resection 
  (b) Example of an HFO in the pre-resection ECoG, wideband and filtered in HFO band (250-500 Hz), and the time-frequency spectrum.
  (c) Firing of SNN neurons indicate the occurrence of an HFO. 
  (d) Pre-resection HFO rates for each recording channel. 
  (e) Electrode position after resection. 
  (f) Example of an HFO in the post-resection ECoG.
  (g) Firing of SNN neurons indicate the occurrence of an HFO. 
  (h) Post-resection HFO rates for each recording channel. 
  In 4 channels (28-29, 29-30, 30-31, 1-2), the HFO rate exceeded 1 HFO/minute. 
  The occurrence of these residual HFOs ``predict'' poor seizure outcome. Indeed, the patient suffered from  recurrent seizures (ILAE 3).}
 \label{fig:HFO_rates}
\end{figure}

For each patient, we counted the number of HFOs detected per electrode channel and divided by the duration of the recording (median 3.5 min, total data duration 58 min) to obtain the HFO rate (Table~\ref{table:patient_characteristics}). We found maximal HFO rates $\geq$ 1 HFO/min in the pre-resection recordings of all 8 patients (8 recordings, median duration 3.9 min, median 6.6 HFO/min, range [1.3 - 45.0] HFO/min). 
In total, the SNN detected 4293 HFOs. 
In the post-resection recordings (8 recordings, median duration 3.0 min), seven of the 8 patients had $<$ 1 HFO/min , i.e. there were no residual HFOs.
Given the absence of residual HFOs, we ``predicted'' good postsurgical seizure outcome in these patients. 
Indeed, these 7 patients achieved seizure freedom after surgery (ILAE 1, median follow-up period after surgery 22 months). 
Only the post-resection recordings of Patient 6 showed an HFO rate $\geq$ 1 HFO/min, which qualifies as residual HFOs. The presence of post-resection HFOs `predicted' poor surgical outcome in this patient who indeed suffered from recurrent seizures (ILAE 3). 
Figure~\ref{fig:HFO_rates} shows the electrode placement, an example of a detected HFO by the SNN, and the HFO rates per recording channel for Patient 6 in both, pre- (Panels a-d) and post-resection recordings (Panels f-j).
Over the group of patients, the prediction accuracy was 100\% (CI [63 100\%]).

\newcommand{\ra}[1]{\renewcommand{\arraystretch}{#1}}
\begin{table}
\ra{1.3}
\caption{
Patient characteristics, HFO rates and postsurgical seizure outcome. We present the maximum HFO rates in the pre- and post-resection ECoG as detected by the Spectrum detector ~\cite{Boran_etal19} and the SNN detector. We ``predict'' seizure outcome for each patient based on whether remaining HFOs were observed in the post-resection recordings or not. We compare the ``prediction'' with the seizure outcome (ILAE).
FCD focal cortical dysplacia; DNET dysembryoplastic neuroepithelial tumors; SW Sturge-Weber, TP True Positive, TN True Negative.}
\label{table:patient_characteristics}
\renewcommand\arraystretch{1.2}
\centering
\resizebox{\textwidth}{!}{\begin{tabular}{@{}cccccccccc@{}}
\toprule
\textbf{Patient} & \textbf{Etiology} & \textbf{Follow-up }& \textbf{Seizure} & \textbf{max HFO rate} & & \textbf{max HFO rate} & &\textbf{Spectrum detector} & \textbf{SNN detector}\\

 &  & \textbf{(months)} & \textbf{outcome} &\textbf{Spectrum detector} & & \textbf{SNN detector} & &\textbf{prediction}  & \textbf{prediction}\\
 
 &  & & \textbf{(ILAE)} & \textbf{pre} &\textbf{post} & \textbf{pre} & \textbf{post} & & \\
 
\hline
1 & DNET           &33 & 1 & 6    & < 1 & 3.4  & < 1  & TN & TN\\
2 & FCD 2b         &24 & 1 & 3.5  & < 1 & 9.7  & < 1  & TN & TN\\
3 & Sturge Weber    &30 & 1 & 2    & < 1 & 1.3  & < 1  & TN & TN\\
4 & Ganglioglioma  &18 & 1 & 7.9  & < 1 & 11.5 & < 1  & TN & TN\\
5 & FCD 2a         &13 & 1 & 13.2 & < 1 & 30   & < 1  & TN & TN\\
6 & Sturge Weber   &20 & 3 & 31.7 & 4.6 & 45   & 13.9 & TP & TP\\
7 & Astrocytoma    &29 & 1 & 1.3  & < 1 & 1.4  & < 1  & TN & TN\\
8 & FCD 2a         &12 & 1 & 22.4 & < 1 & 1.9  & < 1  & TN & TN\\
\bottomrule
\end{tabular}}
\vspace{-0.31cm}
\end{table}

\section{Discussion}
In this study, we simulated a new SNN that detects HFO in intraoperative ECoG. Compared to our previous SNN~\cite{Sharifshazileh_etal20}, we have added an artifact rejection stage to suppress fast transients in the signal. Based on the HFO rates detected in the post-resection ECoG, we were able to ``predict'' the patients' seizure outcome with 100\% accuracy. This is preliminary evidence that the automatically detected HFOs by the SNN may indeed be clinically relevant.

\subsection{Comparison with the Spectrum detector}
The ECoG recordings analysed here have been previously analysed for HFOs with the Spectrum detector~\cite{Boran_etal19}.
The SNN detector in our study performed well while being fully unsupervised, which is an advantage for its possible application in multi-center studies.
In designing the SNN, the aim was not to achieve a one-to-one agreement of the detected HFO events. Rather, the HFO rate threshold of $\geq$ 1 HFO/min in unresected channels was used to ``predict'' seizure recurrence. Both, our simulated SNN and the Spectrum detector reached the same ``prediction'' for each patient of this study (Table~\ref{table:patient_characteristics}).
Even though the SNN prediction of the poor outcome was limited to data from one patient, when not considering only post-resection but also pre-resection recordings, the SNN reached the same qualitative result (HFOs present / HFOs not present) in all 16 recordings.

As a difference to the SNN, the Spectrum detector applied multi-channel information to reject artifacts (Stage III) and, additionally, visual artifact rejection 
These measures were required by the more challenging artifacts in the data recorded with 10 mm electrode distance. As a limitation, in our study we require the integration of densely spaced electrode contacts, low impedance, low noise amplification to enable fully automated HFO detection with the SNN.

\subsection{Clinical considerations}
In our study, we analysed ECoG with a SNN detector that had been previously trained on an independent dataset~\cite{Sharifshazileh_etal20,Fedele_etal17c}. 
The 100\% outcome accuracy speaks for the robustness of automatically detected HFO to predict postoperative seizure freedom.
Our study is limited by the small number of patients, where only one out of the eight patients had recurrent seizures.  Our finding that residual HFO predicted seizure recurrence is in agreement with previous studies showing that incomplete resection of cortical tissue generating HFO correlates with seizure recurrence~\cite{Boran_etal19,Fedele_etal17b,Fedele_etal16,Vant-klooster_etal15a,Vant-klooster_etal17,Vant-klooster_etal15b,Weiss_etal18}. 

\subsection{Future implementation in neuromorphic hardware}
In a previous study, we have designed a neuromorphic device~\cite{Sharifshazileh_etal20} to detect HFO in iEEG during deep sleep~\cite{Fedele_etal17c}. 
That device demonstrated that common pre-processing stages like low-noise amplification, filtering and signal transformation using ADMs can be implemented in the same silicon die alongside a multi-core neuromorphic processor to allow on-line and real-time post-processing of biomedical signals. 
The device is compact, battery-powered and does not interfere with other electronic equipment, which would facilitate its use during surgery.
Seizure outcome prediction on deep sleep iEEG was comparable with the outcome prediction achieved with an HFO detector based on template matching~\cite{Fedele_etal17b}.

In the current study, we have advanced the SNN presented in our earlier study~\cite{Sharifshazileh_etal20} for HFO detection in intraoperative recordings. 
Compared to deep sleep iEEG, the electronic interference from standard machinery in the operating theatre make HFO detection in the intraoperative ECoG more challenging. 
For example, the surgery environment induces fast transients in the wideband ECoG that appear as short oscillations in the HFO frequency range. Hence, for HFO detection in ECoG recordings, we have extended the SNN by adding an artifact rejection stage on top of the core SNN.

Obviously, our new simulated SNN was motivated by a future implementation in neuromorphic hardware~\cite{Borton_etal20}. The parameters of the pre-processing stages were applicable to both iEEG and ECoG and remained unchanged. In the same way, the core SNN performed equally well in the detection of clinically relevant HFO in both iEEG and ECoG with the identical parameter settings.
Also for the artifact rejection stage, all the architecture and parameter decisions have been chosen such that the simulated SNN can easily be implemented in the neuromorphic hardware with only slight adaptations. 

\section{Conclusions}
To detect HFO in intraoperative ECoG, we simulated a SNN extended with an artifact rejection stage to suppress sharp transients. Confirming earlier results, the occurrence of post-resection HFO predicted seizure recurrence. This detector uses spiking neural network computing and thereby is radically different from other detectors. These results provide a further step towards real-time detection of HFO during epilepsy surgery by an SNN implemented in neuromorphic hardware. 

\section*{Acknowledgements}
This project has received funding from Swiss National Science Foundation (SNSF 320030$\_$176222) and from the European Research Council (ERC) under the European Union’s Horizon 2020 research and innovation program grant agreement No 724295.

\section*{Author contributions statement}
JS and GI conceived the experiments, NK and GR treated the patients, KB and MS conducted the experiments and analysed the results. KB and JS wrote the manuscript. All authors critically reviewed the manuscript. 

\section*{Competing interests}
The authors declare no competing interests.

\section*{Data availability}
The ECoG data and HFO markings are freely available at \url{https://gin.g-node.org/USZ_NCH/Intraoperative_ECoG_HFO.}
The code for the SNN detector is available at the GitHub repository \url{https://github.com/kburel/SNN_HFO_ECoG}

\bibliography{biblioncs.bib}

\end{document}